\documentclass[12pt]{article}
\usepackage{graphicx}

\def\a{\alpha}
\def\b{\beta}
\def\c{\chi}

\def\e{\epsilon}
\def\f{\phi}

\def\l{\lambda}
\def\m{\mu}

\def\p{\pi}

\def\D{\Delta}

\def\S{\Sigma}

\def\vf{\varphi}

\def\beq{\begin{equation}}
\def\eeq{\end{equation}}
\def\bea{\begin{eqnarray}}
\def\eea{\end{eqnarray}}

\def\NO{\nonumber}

\def\pl#1#2#3{Phys.~Lett.~{\bf B {#1}} ({#2}) #3}
\def\np#1#2#3{Nucl.~Phys.~{\bf B {#1}} ({#2}) #3}
\def\prl#1#2#3{Phys.~Rev.~Lett.~{\bf #1} ({#2}) #3}
\def\pr#1#2#3{Phys.~Rev.~{\bf D {#1}} ({#2}) #3}

\def\prep#1#2#3{Phys.~Rep.~{\bf {#1}C} ({#2}) #3}

\newcommand{\mpl}{M_{\rm P}}
\newcommand{\VEV}[1]{\langle #1 \rangle}

\newcommand\gev{\,\mbox{GeV}}

\renewcommand{\(}{\left(}
\renewcommand{\)}{\right)}
\renewcommand{\[}{\left[}
\renewcommand{\]}{\right]}

\begin{document}
\date{\mbox{ }}
\title{{\normalsize DESY 00-082\hfill\mbox{}\\
June 2000\hfill\mbox{}}\\
\vspace{2cm} \textbf{Inflation and Supersymmetry Breaking}\\
[8mm]}
\author{Wilfried~Buchm\"uller, Laura Covi and David~Del\'epine \\
\textit{Deutsches Elektronen-Synchrotron DESY, Hamburg, Germany}}
\maketitle

\thispagestyle{empty}

\begin{abstract}
\noindent
We study the connection between inflation and supersymmetry breaking in
the context of an O'Raifeartaigh model which can account for both 
hybrid inflation and a true vacuum where supersymmetry is spontaneously
broken. For a weakly coupled inflaton field, the dynamics during the
inflationary phase can be determined by the supersymmetry breaking scale
$M_S\sim 10^{10}\gev$, even if $H_I \gg m_{3/2}$. The spectrum of density 
fluctuations is then almost 
scale invariant, with a spectral index $n-1={\cal O}(M_G^2/\mpl^2)$. The mass 
parameter $M_G$ of the O'Raifeartaigh model is determined by the COBE 
normalization for the cosmic microwave background to be the grand unification 
scale, $M_G \sim 10^{16}\gev$.
\end{abstract}

\newpage

It is well known that an inflationary phase in the early history of the
universe can explain its present flatness, isotropy and homogeneity 
\cite{kt90}. 
From the COBE measurement of the cosmic microwave background (CMB)
anisotropy it has soon been realized that the scale of inflation has to be 
lower than the Planck scale, but much larger than the electroweak scale. 
It is therefore clear that supersymmetry may play an important role for 
inflation. Many models have been proposed describing an inflationary phase
in the context of globally supersymmetric theories \cite{lr99}.
But since supersymmetry is not exact in nature, we know that globally 
supersymmetric models can give us only an approximate description of the real 
world. Supergravity corrections can strongly 
affect the inflationary phase \cite{cll94}, and a variety of models have 
been constructed in a general supergravity framework \cite{lr99}. 
However, it is often thought that, when the scale of inflation is much larger 
than the supersymmetry breaking scale, a globally supersymmetric model is 
sufficiently accurate to describe the inflationary phase as well as the 
reheating process.

As we shall see, this is not the case. The goal of 
this paper is to study explicitly a model containing both, supersymmetry 
breaking in the true vacuum and during the inflationary phase
\footnote{For related earlier work, see \cite{sbinfl}}. 
On the one hand, in such a case the supersymmetry breaking sector is 
influenced by the inflaton dynamics, and the true vacuum is reached only at 
the end of inflation. On the other hand, also the inflationary potential
is modified by the presence of the supersymmetry breaking sector. This 
modification turns out to be very important in the case of a weakly coupled
inflaton field.

In the following we first describe the model, which 
combines a Fayet term \cite{fa75} for global symmetry breaking with a Polonyi
term \cite{po77} for supersymmetry breaking, leading to a particular  
O'Raifeartaigh model \cite{or75}. We then analyze the model without and with
supergravity corrections. Finally, we briefly discuss the moduli problem and 
the reheating process.\\

\noindent\textbf{The model}\\

The usual hybrid inflation scenario is based on a superpotential
of the type
\beq
W_G = \lambda T \(M_G^2 - \Sigma^2\)\;,
\label{W-infl}
\eeq
where $T$ and $\Sigma$ are chiral superfields.
Such a potential is well-known and has initially been used \cite{fa75} 
to break global or local symmetries in supersymmetric theories. Apart from 
being the simplest choice giving hybrid inflation 
\cite{cll94},\cite{li91}-\cite{ahk00}, it has also the 
advantage of avoiding large supergravity corrections in the case of a 
canonical K\"ahler potential. This is due to the fact that the superpotential 
is linear in the inflaton field $T$. In (\ref{W-infl}) higher powers of
$T$ are forbidden by R-invariance. 

To the potential (\ref{W-infl}) we add, as supersymmetry breaking part, the 
Polonyi potential
\beq
W_S = M_S^2 (\beta + S)\;,
\label{W-P}
\eeq
with $\beta = (2-\sqrt{3}) \mpl$, which allows for supersymmetry breaking 
in the true vacuum where $\VEV{S} = (\sqrt{3}-1)\mpl$. $M_S$ is the scale of 
supersymmetry breaking yielding the gravitino mass 
$m_{3/2} = M_S^2/\mpl \exp (2-\sqrt{3})$ and 
$\mpl=(8\p G_N)^{-1/2}=2.4\cdot 10^{18}\gev$. 
For numerical estimates we shall   
use $m_{3/2}\simeq 100$~GeV, which corresponds to the
supersymmetry breaking scale $M_S \simeq 1.4\cdot 10^{10}$~GeV. 

In the case of global supersymmetry the scalar Polonyi potential is flat,
\beq
V_S = \left|{\partial W_S\over \partial S}\right|^2 = M_S^4\;.
\eeq
This has motivated early attempts to identify the Polonyi field $S$ with the 
inflaton field \cite{os83}. However, supergravity corrections turn out to be 
too large and spoil the flatness of the potential.

The constant $\b$ in the Polonyi potential (\ref{W-P}) is only relevant in the
supergravity framework. It breaks R-invariance, and it is adjusted to have 
vanishing cosmological constant. As we shall see, this constant may play an
important role during inflation. Our conclusions will generally apply to any 
supersymmetry breaking effective potential where the constant and the linear
term dominate in an expansion in powers of $S$.

Combining the two superpotentials (\ref{W-infl}) and (\ref{W-P}) we arrive at
\beq
W = W_G + W_S = \lambda T \(M_G^2-\Sigma^2\)+ M_S^2 \(\beta + S\)\;.
\label{W-1}
\eeq
Further, we choose the canonical K\"ahler potential for the fields
$T, \Sigma $ and $S$. Note, that the superpotential (\ref{W-1}) is  
a particular O'Raifeartaigh model. This becomes apparent after a change 
of variables. Defining
\beq
\Phi = {\xi\, S\over \sqrt{1+\xi^2}} + {T\over \sqrt{1+\xi^2}}\;,\quad
\Psi = {S\over \sqrt{1+\xi^2}} - {\xi\, T\over \sqrt{1+\xi^2}}\;,
\eeq
with $\xi = M_S^2/(\lambda M_G^2)$, and
\beq
\lambda_1 = {\lambda\over \sqrt{1+\xi^2}}\;,\quad
\lambda_2 = {\lambda\, \xi\over \sqrt{1+\xi^2}}\;,\quad
M = M_G \sqrt{1+\xi^2}\;,
\eeq
one obtains
\beq
W = \lambda_1 \Phi (M^2 -\Sigma^2) + \lambda_2 \Psi \Sigma^2
+ M_S^2 \beta\;.
\label{W-2}
\eeq
This is the more familiar form of an O'Raifeartaigh model \cite{or75}.
The two superpotentials (\ref{W-1}) and (\ref{W-2}) are equivalent, and in the 
following we will use one or the other according to our convenience. As
we shall see, successful inflation requires $\xi$ to be very small, so that
effectively $T\simeq \Phi$, $S\simeq \Psi$ and $M\simeq M_G$.\\

\noindent\textbf{Hybrid inflation}\\

For global supersymmetry the scalar potential reads
\beq
V_G + V_S = \l^2 |M_G^2-\S^2|^2 + 4\l^2 |T|^2 |\S|^2 + M_S^4\;,
\eeq
and the corresponding ground state is given by 
\beq
\VEV{T} = 0\;,\quad \VEV{\Sigma} = M_G\;, 
\eeq
while $\VEV{S}$ is undetermined. Hence, the supersymmetry breaking sector 
decouples, and an inflationary phase can take place as in ordinary hybrid 
inflation, starting with a large value of $T$. The field $\Sigma$ is then 
pushed to the origin by a large mass term and the potential is perfectly flat 
along $T$. A small curvature needed for the `slow roll' is generated by the 
quantum corrections due to the loops of the $\Sigma$ particles \cite{dss94}, 
which are non-vanishing since supersymmetry is broken by 
$F_T=\partial W_G/\partial T \neq 0$. The corresponding one-loop correction 
to the scalar potential reads,
\beq
\D V_G = {\l^4 M_G^4 \over 8\pi^2}
\(\ln (2\lambda^2 \phi^2/\m^2) + O(M^4_G/\phi^4)\)\;,
\label{V-CW}
\eeq 
where $\phi$ is the real part of the complex scalar field $T$ and $\m$ is a
renormalization scale.

Inflation ends at $\phi_c\simeq M_G$, where the mass of $\Sigma$ becomes
negative and the field acquires a non-vanishing expectation value. 
For $M_S \ll \l^{1/2}M_G$, the potential (\ref{V-CW}) satisfies the
slow-roll conditions \cite{lr99} down to $\phi_c$,
\bea
\epsilon &=& {\mpl^2\over 2} \({V'\over V}\)^2 
= {\lambda^4\over 32 \pi^4} {\mpl^2\over \phi^2} \ll 1\;, \\
\eta &=& \mpl^2 {V''\over V} = 
- {\lambda^2\over 4\pi^2} {\mpl^2\over \phi^2}\ ,\quad |\eta| \ll 1\;,
\eea
as long as $\lambda$ is of order $M_G/\mpl$.

The number of e-folds between the inflaton field value $\phi$ and the
end of inflation at $\phi_c$ is given by
\beq
N(\phi) = \int_{t(\phi_c)}^{t(\phi)} H_I d t 
= \int_{\phi_c}^{\phi} {V\over V'} d\phi
= {2\pi^2\over \lambda^2} {\phi^2 - \phi^2_c\over \mpl^2}\;,
\eeq
where $t$ denotes time and
$H_I \simeq \sqrt{V_G/(3\mpl^2)} \simeq \lambda M^2_G/(\sqrt{3}\mpl) $ in 
the slow-roll approximation. 

Adiabatic density perturbations originate as vacuum fluctuations during 
inflation. The COBE normalization \cite{COBE} then gives
\beq
\delta_H \equiv {1\over \sqrt{75} \pi \mpl^3} 
{V_*^{3/2}\over |V_*'|} =  {4\pi\over \sqrt{75}} 
{M_G^2 \f_*\over \lambda \mpl^3} =
1.94 \times 10^{-5}\; ,
\label{cobe-norm}
\eeq
where the $*$ indicates that the potential and its derivative are evaluated 
at the epoch of horizon exit for the comoving scale 
$k_*\simeq 10 H_0$ \cite{COBE}. 
Defining $N_*$ as the number of e-folds at that epoch, the corresponding 
inflaton value is given by 
$\phi_*/\mpl \simeq \sqrt{\l^2 N_*/(2 \pi^2)+\phi^2_c/\mpl^2}$. We thus
obtain the relation
\beq\label{mgscale}
{M^2_G\over \mpl^2}
\left(N_* + {2\pi^2\over \l^2}{\phi_c^2\over \mpl^2}\right)^{1/2} 
\simeq 5.9 \cdot 10^{-5}\; ,
\eeq
where the number of e-folds is $N_* \simeq 50$.

\begin{figure}
\centering 
\includegraphics[scale=0.6]{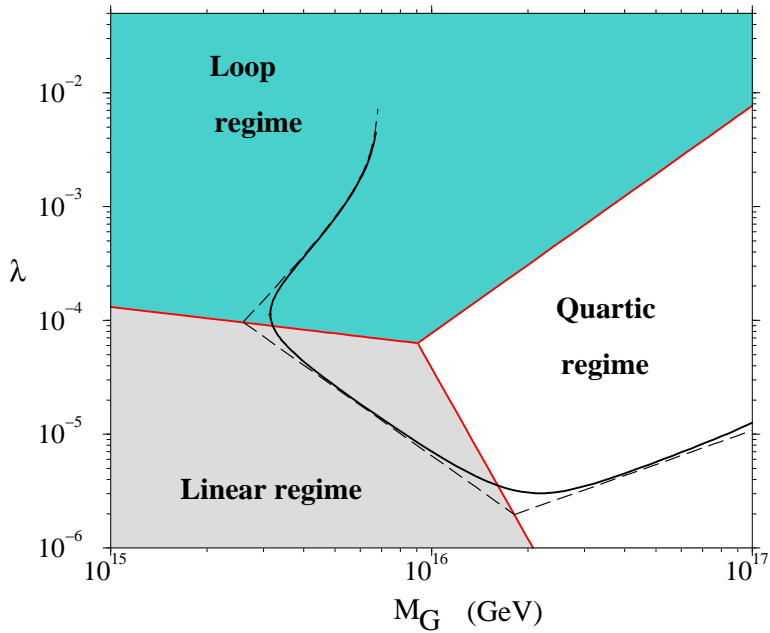}
\caption{{\it The three regimes of hybrid inflation: the loop regime, the
linear regime and the quartic regime. The COBE normalization defines a curve
(full line) in the $\l - M_G$ - plane. The dashed lines are obtained by
assuming approximate forms of the inflaton potential as discussed in the text.
}}
\end{figure}

From eq.~(\ref{mgscale}) one reads off that a consistent picture is obtained
for $M_G/\mpl \sim \l \sim 10^{-3}$, with $H_I \sim 10^{-8}\mpl$.
The exact relation between $\l$ and $M_G$ imposed by the COBE constraint is
shown in fig.~(1). It is interesting that $M_G$ is naturally of order the 
grand unification scale. Moreover, due to the smallness of $\l$ the observable
number of e-folds corresponds to inflaton fields $\phi$ close to the critical
value $\phi_c \simeq M_G \ll \mpl$.\\

\noindent\textbf{Supergravity corrections}\\

Let us now consider the effect of the supersymmetry breaking Polonyi 
potential and of corrections suppressed by powers of $1/\mpl$. The 
supergravity scalar potential reads 
\beq
V = e^{K/\mpl^2} 
\[\left|{\partial W \over \partial z_i} + {z_i^* W\over \mpl^2} 
\right|^2 - 3 
{|W|^2 \over \mpl^2} \]\;,
\eeq
where the sum extends over all fields $z_i$, and $K$ is chosen to be the 
canonical K\"ahler potential, $K = \sum_i |z_i|^2 $.

The additional non-renormalizable terms modify slightly the vacuum expectation
values of the fields $T$ and $\Sigma$ and give a large expectation value to the
Polonyi field $S$. Since the corrections to the derivatives of the 
superpotential are always proportional to $\VEV{W}/\mpl^2\ll 1$, it is 
possible to expand the potential in powers of $M_S\over \mpl$. This 
yields for the first corrections to the vacuum expectation values,
\bea
\VEV{\Sigma} &=& M_G \[ 1 + \frac{2-\sqrt{3}}{4}
\frac{M_S^4}{\lambda^2 \mpl^2 M_G^2} + O\({M_S^3\over \mpl^3}\) \]\;,\\
\VEV{T} &=& {1\over 2\lambda} {M_S^2\over \mpl} + O\({M_S^3\over \mpl^2}\)\;.
\eea
Also the value of $\b$, which is adjusted to have vanishing cosmological 
constant, and the vacuum expectation value of $S$ acquire corrections, 
\bea
\VEV{S} &=& (\sqrt{3}-1) \mpl -\frac{9-4\sqrt{3}}{24}
\frac{M_S^4}{\lambda^2 \mpl^3}\;,\\
\beta &=& (2-\sqrt{3}) \mpl -\frac{\sqrt{3}}{6}\frac{M_G^2}{\mpl}\;. 
\eea
Clearly, all corrections ${\cal O}(1/\mpl^n)$ to the vacuum expectation values
are very small. One may therefore be tempted to think that also during the
inflationary phase supergravity corrections are practically negligible.
This, however, is not the case.

During the inflationary phase $\Sigma $ is driven to zero by the large value
of $T$. The potential is then most easily computed in the basis $\Phi, \Psi$. 
Neglecting the one-loop correction, one has
\bea
V &=& \l^2 M_G^4 \left(1+\xi^2- {2\sqrt{2}\xi \b \vf\over \mpl^2}  
-{\xi^2\b^2(\vf^2+\chi^2)\over \mpl^4}-{\xi\b\vf^3\over \sqrt{2}\mpl^4}
-{\sqrt{2}\xi\b\vf\chi^2\over \mpl^4}\right.\NO\\
&&\hspace{2cm}\left.
+ {(\vf^2+\c^2)^2\over 8 \mpl^4} 
+ {|\Psi|^2\over \mpl^2}\left(1+{\vf^2+\chi^2\over 2\mpl^2}\right) 
+ \ldots \right)\;,
\label{V-sugra}
\eea
where $\xi = M^2_S/ (\l M^2_G) \ll 1$ and $\Phi = (\vf+i\c)/\sqrt{2}$. Here we
have neglected terms which are small for values of $\Phi$ in the range  
$1 > |\Phi| \gg \xi$. Note, that the potential for $\Phi$ at $\Psi=0$ is just 
the Polonyi potential, but with the `wrong' constant $\xi\b$, i.e. while the 
supersymmetry breaking scale during inflation is given by $\lambda^{1/2} M_G$,
the constant is still related to the supersymmetry breaking scale $M_S$ in the 
true vacuum. The hierarchy between the two scales is exactly what makes the 
potential flat enough, contrary to the simple expectation for the Polonyi 
potential with only one scale of supersymmetry breaking. This hierarchy also
implies that in the potential (\ref{V-sugra}) the term linear in $\vf$ is
larger than the $\vf$ mass term, which is suppressed by an additional
power of $\xi$.

We remark that in the case of a charged inflaton field, or in general when 
the superpotential contains only second and higher powers of the inflaton
field, no linear term is generated by the supersymmetry breaking sector.
Then the first supergravity correction is a mass term and
inflation can be realized even with $H \simeq m_{3/2}$, if the inflaton mass 
is sufficiently suppressed either by cancelations or by the running mass 
mechanism \cite{run}.

The minimum of the potential (\ref{V-sugra}) with respect to $\Psi$ and 
$\chi$ lies at the
origin, but it is very flat. However, for initial values 
$\vf = {\cal O}(\mpl)$ one can have $m_{\Psi}, m_{\chi} > H$, which may be
sufficient to drive $\Psi$ and $\chi$ to the origin before the beginning
of inflation. In the following we shall assume $\Psi\simeq \chi\simeq 0$
as initial conditions. 

Comparing (\ref{V-sugra}) with (\ref{V-CW}) it is clear that the 
standard hybrid inflation scenario may be significantly modified depending
on the values of $\l$ and $M_G$. The one-loop radiative corrections
dominate over the linear term in (\ref{V-sugra}) for
$\l^2 / (2\pi)^2 \geq \xi \b \vf/ (\sqrt2 \mpl^2)$.
Substituting $\b/\mpl = 2-\sqrt{3}$ and using $\vf > \vf_c \geq M_G$, one
obtains the lower bound on $\l$,
\beq
\lambda > 3 \({M_S^2\over M_G \mpl}\)^{1/3}\;.
\eeq 
For the hybrid inflation value $M_G \simeq 3\cdot 10^{-3} \mpl$, this yields
$\lambda > 0.7\cdot 10^{-4}$. As discussed above, hybrid inflation takes
place in the vicinity of $\vf_c$. For couplings
$\l$ above the lower bound the one-loop radiative corrections also dominate
over the supergravity induced quartic term in (\ref{V-sugra}). Hence,
for $M_G \simeq 3 \cdot 10^{-3}\mpl$ and couplings in the range
\beq
0.7\cdot 10^{-4} < \l < 6\cdot 10^{-3}
\eeq
the standard hybrid inflation scenario is only weakly affected by supergravity 
corrections.\\

\noindent\textbf{Scale invariant inflation}\\

Consider now the case of small couplings,
\beq\label{uplam}
\lambda < 3 \({M_S^2\over M_G \mpl}\)^{1/3}\;,
\eeq 
for which the linear term in (\ref{V-sugra}) dominates over the one-loop 
radiative corrections (cf.~fig.~1). From the COBE normalization 
(\ref{cobe-norm}) one then obtains, independently of $N_*$,
\beq\label{delh}
\delta_H = {1\over \sqrt{75} \pi \mpl^3} 
{V_*^{3/2}\over |V_*'|} ={1\over 2\sqrt{150}\pi}{M_S^2 \over \xi^2 \beta \mpl} 
\simeq 1.9\cdot 10^{-5}\;.
\eeq
Fixing $M_S \simeq 1.4\cdot 10^{10}\gev$,
this implies $\xi \equiv M^2_S/(\l M_G^2) \simeq 5\cdot 10^{-7}$. Note, that 
$\xi$ is the ratio of the gravitino masses in the true vacuum and in the
inflationary phase. Since $\xi \ll 1$, a huge number of e-folds 
is generated near $\vf_c = -M_G$, 
\beq
N(\phi) = - {M_G\over 2\sqrt{2}\,\xi\beta}\, {\vf - \vf_c\over\vf_c}\;,
\eeq
which include the cosmologically relevant scales with $N\simeq 50$. The
linear term dominates over the quartic term in (\ref{V-sugra}) if
$|\vf^3|/(2\mpl^2)< 2\sqrt{2}\xi\b$. Together with (\ref{delh}), this yields
an upper bound on $M_G$. Similarly, a lower bound on $M_G$ follows from
(\ref{uplam}). Inserting numerical values for $\xi$ and $M_S$ one finds
that the linear term dominates in the range
\beq
2\cdot 10^{15}\mbox{GeV} < M_G < 2\cdot 10^{16}\mbox{GeV}\;.
\eeq

The slow-roll conditions are clearly satisfied for $\vf \simeq \vf_c$, 
\bea
\e &=& {4 \xi^2 \b^2\over \mpl^2}
\(1- {\vf^3\over 2\sqrt{2} \xi\b \mpl^2} + \ldots\)\ \ll 1\;, \\
\eta &=& {3\over 2} {\vf^2\over \mpl^2} + \ldots\ \ll 1\;.
\eea
Here we have kept the quartic supergravity correction to the linear term
in (\ref{V-sugra}), which affects the spectral index,
\beq
n -1 \simeq 3 {\vf_*^2\over \mpl^2} \leq  2.4 \cdot 10^{-4}\;.\label{n-1}
\eeq
An inflationary phase dominated by a linear term is very interesting, since 
it gives a {\it scale invariant spectrum} to high accuracy. For standard
hybrid inflation, on the contrary, one has $n\simeq 0.98$ \cite{dss94}. 
Future satellite 
experiments may eventually be able to distinguish between these two versions
of hybrid inflation.

The linear term in the potential breaks the symmetry $\vf \rightarrow -\vf$.
(cf.~fig.~2). For negative $\vf$ hybrid inflation can take place, as discussed
above. The potential has a Polonyi-type minimum at 
$\vf_{min} \simeq (4\sqrt{2} \xi \beta/\mpl)^{1/3}
\mpl \simeq 9\cdot 10^{-3} \mpl$. For positive initial condition hybrid
inflation can take place as long as $M_G > \vf_{min}$. Otherwise, the inflaton
field is trapped at $\vf_{min}$. Inflation then continues in this
metastable state and has to terminate in a different way.  

Let us finally turn to the case where the quartic term dominates during
inflation, a possibility already considered in \cite{quartic}. This occurs
for $M_G > 2\cdot 10^{16}$~GeV. The slow-roll conditions,
\bea
\e &=& {\vf^6\over 4\mpl^6} 
\(1- {2\sqrt{2} \xi\b \mpl^2\over \vf^3} + ...\) \ll 1\;, \\
\eta &=& {3\over 2} {\vf^2\over \mpl^2} + \ldots \ll 1 \;,
\eea
are satisfied for field values small compared to $\mpl$. The number of e-folds
is given by
\beq
N(\vf) = \int_{\vf_c}^{\vf} d\vf {2 \mpl^2\over \vf^3}
= {\mpl^2\over \vf_c^2}-{\mpl^2\over\vf^2}\;. 
\eeq
For $\vf_c < 10^{-1} \mpl$ the cosmologically relevant scales 
again correspond to $\vf_* \simeq \vf_c$.

The COBE normalization determines $\l$ as function of $M_G$ (cf.~fig.~1),
with $\l > 3\cdot 10^{-6}$. For the spectral index one obtains 
\beq
n -1 \simeq 3 {\vf_*^2\over \mpl^2} \geq  2.4 \cdot 10^{-4}\;.
\eeq

\begin{figure}[t]
\centering 
\includegraphics[scale=0.6]{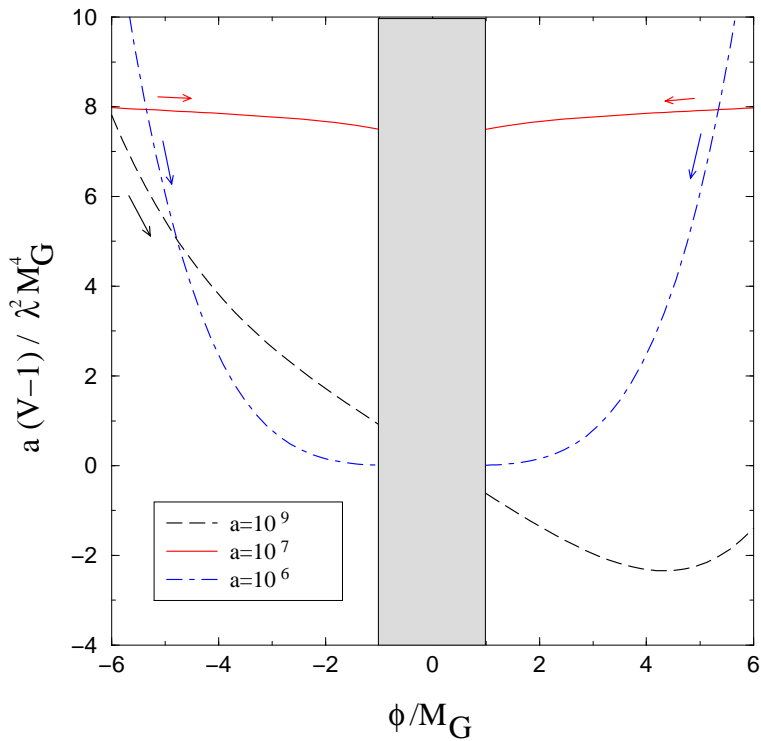}
\caption{{\it The full inflaton potential for three sets of parameters
$(\l,M_G)$: $(10^{-3},5\cdot 10^{15}\gev)$,  
$(1.5\cdot 10^{-5},5\cdot 10^{15}\gev)$ 
and $(4\cdot 10^{-6},4\cdot 10^{16}\gev)$, 
corresponding to the loop regime (dotted line), the linear regime (dashed line)
and the quartic regime (dashed-dotted line), respectively. The 
flattest potential is obtained in the linear regime (Note the different 
rescaling factors!).
The arrows indicate the directions of possible inflationary phases.}}
\end{figure}

The three regimes of hybrid inflation, the loop regime, the linear regime
and the quartic regime, are summarized in fig.~1. The COBE normalization
defines a curve in the $\l-M_G$ - plane. The dashed lines are obtained by
assuming that a single term dominates the derivative of the supergravity
potential. The full line is based on the full potential. Increasing 
(decreasing) the scale of supersymmetry breaking $M_S$ shifts the curve
in the linear regime, as well as the boundaries,
to larger (smaller) values of $\l$. The inflaton potentials in the three
regimes are compared in fig.~2. Note, that the flattest potential corresponds
to the linear regime.\\

\noindent\textbf{Moduli problem and reheating}\\

At the end of the inflationary period the field $\S$ has to change from
$0$ to $M_G$, the field $T$ from $M_G$ to $M_S^2/(2\l\mpl)$ and the field
$S$ from $0$ to $(\sqrt{3}-1)\mpl$. As in the usual Polonyi model S acquires
a small mass $m_S \sim m_{3/2}$ in the true vacuum, like the standard model
fields which have only gravitational interactions with $S$. The late decays 
of $S$ are then incompatible with nucleosynthesis, which is the so-called
cosmological moduli problem \cite{Polonyi-pbm}.

Several ways have been proposed to circumvent the moduli problem. 
For instance, it does not occur if the amplitude of the moduli field is
reduced via an effective mass term during the evolution to the true vacuum  
\cite{linde96}. This can be implemented in the present model by adding the
following non-renormalizable term to the superpotential, 
\beq
W_{M}\equiv \frac{\a}{\mpl}S^2T^2\; .
\label{W-PP}
\eeq
This interaction is negligible during inflation, where $S \ll \mpl$, and 
modifies only slightly the expectation values in the true vacuum.

However, at the end of inflation, $T\simeq M_G$ and $S$ acquires the mass
$m_S = 2 \a M_G^2/\mpl$. The amplitude of the S field oscillations is then 
sufficiently damped for $m_S \gg H_I=\l M_G^2/(\sqrt{3}\mpl)$ \cite{linde96}.
This is the case for $\a\gg \l$, which can be easily satisfied.

The interaction (\ref{W-PP}) also induces a large mass for the T field, 
$m_T \sim \a \mpl \gg \l M_G$, when $S$ approaches its minimum. 
$T$ then decays rapidly to other particles. The computation of the 
corresponding reheating temperature is not straightforward, since it depends 
on the dynamics of the field $S$. A detailed analysis of this process, 
including the thermal and non-thermal production of gravitinos, is in 
progress. A rough estimate, providing a lower bound on the reheating 
temperature, can be obtained by considering the decay of $\Sigma$ into quarks.
Supergravity always induces the non-renormalizable couplings
\beq
{\cal L} = Y_q Q H q {|\Sigma |^2 \over \mpl^2}\;,
\eeq
where $Y_q$ is the quark Yukawa coupling, $Q$ ($q$) the quark doublet 
(singlet) and $H$ the corresponding Higgs doublet. From the top-quark 
contribution alone, one obtains a reheating temperature $T_{R} \sim 10^6 \gev$
\cite{kt90}. Clearly, this simple picture may be strongly modified by 
additional interactions.

Finally, let us comment on the possible production of topological 
defects at the end of inflation. The potential (\ref{W-1}) has a 
$Z_2$ symmetry with respect to the field $\Sigma$, that would give 
rise to domain walls at the end of inflation
\cite{dw}. In order to avoid them, it is sufficient, either to
consider higher order non-renormalizable terms breaking the $Z_2$
symmetry or, like in many formulations of hybrid inflation, give to the
$\Sigma $ field a charge under a gauge group and substitute
$\Sigma^2 $ by $\Sigma \bar\Sigma$ or $ Tr (\Sigma^2)$,
depending on the gauge group representation. In the last case, at the
end of inflation other topological defects could arise, e.g. strings,
and could give a non-negligible contribution to the density perturbations
\cite{chm98}.\\

\noindent\textbf{Conclusions}\\

We have studied in detail the connection between inflation and supersymmetry
breaking in the context of an O'Raifeartaigh model, which can account
for both a hybrid inflationary phase and a true vacuum where supersymmetry
is spontaneously broken. Crucial ingredients of the model are two 
contributions to the superpotential: a term linear in the inflaton field and
a constant which is required by the nearly vanishing cosmological constant in 
the true vacuum. This constant generates a linear and higher order terms
in the inflaton field. This does not spoil the flatness of the inflaton
potential since the energy scale during inflation turns out to be large
compared to the scale of supersymmetry breaking. For the same reason the
linear term in the inflaton potential dominates over the mass term. 

The dynamics during the inflationary phase depends on the size of the Yukawa 
coupling in the O'Raifeartaigh model. For $\l > 10^{-4}$, the usual picture
of hybrid inflation driven by loop corrections
applies. However, for smaller coupling $\l < 10^{-4}$, and 
$M_G < 2\times 10^{16} \gev$,
the linear term in the effective potential dominates the evolution of the
inflaton field. As a consequence, the spectrum of fluctuations is almost
scale invariant. The deviation of the spectral index from one is determined
by the mass parameter $M_G$ of the O'Raifeartaigh model,
$n-1={\cal O}(M_G^2/\mpl^2)$. It is remarkable that for $M_S \sim 10^{10}\gev$
the COBE normalization for the cosmic microwave
background determines $M_G$ to be the unification scale, 
$M_G \sim 10^{16}\gev$. \\

\noindent
We would like to thank D.~H.~Lyth for clarifying comments.

\end{document}